\documentclass[12pt]{iopart}

\pdfoutput=1 

\usepackage[left=1.5cm, right=1.5cm, top=1.785cm, bottom=2.0cm]{geometry}
\usepackage{balance}
\usepackage{mathptmx}
\usepackage{graphicx} 
\usepackage{float}
\usepackage{fancyhdr}
\usepackage{fnpos}
\usepackage{array}
\usepackage[T1]{fontenc}
\usepackage{hyperref}

\begin{document}
\title[Electrode insulation]{A Simple Electrode Insulation and Channel Fabrication Technique for High-Electric Field Microfluidics}

\author{Gaurav Anand$^1$, Samira Safaripour$^1$, Jaynie Tercovich$^1$, Jenna Capozzi$^1$, Mark Griffin$^1$, Nathan Schin$^1$, Nicholas Mirra$^1$, Craig Snoeyink$^1$}

\address{$^1$Department of Mechanical and Aerospace Engineering, \\
State University of New York at Buffalo, Buffalo, NY, USA}
\ead{gauravan@buffalo.edu,craigsno@buffalo.edu}

\begin{abstract}
A simple and robust electrode insulation technique that can withstand a voltage as high as $\mathrm{1000~V}$, which is equivalent to an electric field strength of $\sim 1MV/m$ across a $\mathrm{10~\mu m}$ channel filled with an electrolyte of conductivity $\sim 0.1~S/m$, i.e., higher than sea water's conductivity, is introduced. A multi-dielectric layers approach is adopted to fabricate the blocked electrodes, which helps reduce the number of material defects. Dielectric insulation with an exceptional breakdown electric field strength for an electrolyte confined between electrodes can have a wide range of applications in microfluidics, like high electric field strength-based dielectrophoresis. The voltage-current characteristics are studied for various concentrations of sodium chloride solution to estimate the insulation strength of the proposed materials, and the breakdown strength is calculated at the point where the electrical insulation failed. A detailed adhesion technique is also demonstrated, which will reduce the ambiguity around the fabrication of a sealed channel using SU-8.

\end{abstract}

\vspace{2pc}

\normalsize

\section{Introduction}

High electric fields in the absence of active electrodes are common in microfluidic applications. Electrowetting on dielectrics is a mature field with commercial applications in variable focusing lenses, programmable reflectivity, adhesion modification, and microfluidic lab-on-a-chips.\cite{nelson_droplet_2012,li_current_2020,welters_fast_1998} Dielectrophoresis has been used to manipulate particles, positioning and rotating them based on their electrical properties.\cite{pesch_review_2021,castellanos_electrohydrodynamics_2003} Electrothermal pumping can be used to develop strong flows in microfluidic channels both for pumping and mixing.\cite{williams_electrothermal_2015,zhang_study_2013} Dielectrophoresis and electrothermal pumping can be combined to produce electrokinetic vortices which collect and sort particles.\cite{kumar_experiments_2008,salari_ac_2019} Finally, high electric fields are used to fabricate and manipulate vesicles.\cite{yamamoto_stability_2010} 

The electric fields in these applications can be quite high, near the dielectric breakdown strength of the insulators, which is a result of the small electrode gaps that exist in microfluidic applications. Field strengths in electrowetting and dielectrophoresis can approach $\mathrm{4x10^{7}~V/m}$, near the breakdown strength of water.\cite{nelson_droplet_2012} In this regime, ions in solution can penetrate the dielectric and initiate breakdown well before the quoted dielectric strength of the material.\cite{raj_ion_2009} The necessity of high electric field and low operational voltage in electrowetting is addressed by using very thin dielectric layers of thickness of a few hundreds of nanometers. Though a comprehensive study of different dielectric materials and their thickness to be used in electrowetting based applications is readily available, a similar study for dielectrophoresis based applications is missing. Also, the dependence on both geometry and medium make it difficult to generalize published dielectric strengths that don't take these into account. Therefore, we present a simple insulation fabricating technique for dielectrophoresis based applications, characterized using parallel plate electrodes, that will prove to be a significant step in advancement of various new applications which uses a very high electric field. The presented fabrication method will add to the existing literature as a reliable method to produce microfluidic chips that has a very high dielectric breakdown strength that will help improving the dielectrophoretic separation efficiency in biological and non-biological systems.\cite{nakano_protein_2013,hayes_dielectrophoresis_2020,pesch_review_2021}

To prevent faradaic reactions at the electrodes, a wide range of insulating materials have been developed using a variety of coating techniques. Coating materials like parafilm or food wrap film have been successfully used at voltages as high as 300 V.\cite{wang_electrowetting--dielectric_2020, chips_two-for-one_nodate} While very economical, there is no clear way to fabricate channels with such films limiting their use to digital/droplet-based microfluidics. Dielectric layers like $\mathrm{SiO_2}$ that are grown, usually through some variety of vapor deposition techniques, require specialized machines and can be costly. These layers can have dielectric breakdown strengths of well above $\mathrm{350~MV/m}$,\cite{hong_fast_2015,chang_driving_2010} though breakdown voltages of as low as $\mathrm{70~V}$ have been reported for thin layers,\cite{lin_low_2010} highlighting the difficulty in interpreting dielectric breakdown results.

A flexible and economical method that lies in the middle ground between vapor deposition and applied bulk films is the spin-coating of dielectric layers. A few common dielectrics used with spin coating include polydimethylsiloxane (PDMS), SU-8, Teflon AF, Parylene, and Poly(methyl methacrylate) (PMMA). PDMS, though easily accessible and easy to fabricate, is not suitable for electric insulation purposes because it has a high permeability that makes it chemically nonresistant.\cite{liang_robust_2018} Hydrophobic coatings like Teflon AF, Cytop, and Parylene C have been endorsed as perfect insulators by many researchers due to their extraordinary dielectric properties.\cite{lin_low_2010,papageorgiou_superior_2012,zhou_experimental_2019} But the inaccessibility and high cost have thwarted their full potential to be used in any real-life applications. SU-8 and PMMA are highly accessible materials, have excellent dielectric properties, and are highly chemical resistant, which makes them ideal candidates for applications involving electrolytes.\cite{zhang_sacrificial_2017,lima_glasssu-8_2013} Thus, this report explores the practicality of SU-8 and PMMA to be used as an insulator in high electric field applications and will minimize the effect of electrode geometry by utilizing a parallel plate electrode configuration. 

The electrical properties of PMMA\cite{zhang_optical_2015} and SU-8\cite{melai_electrical_2009} as an isolated layer for air have been studied before but have never been investigated in homogeneous, multiple-layer systems with electrolytes. We opted multiple-layers approach for electrolytes because there are clear advantages of having a multi-layer insulator as it reduces defects by overlapping with another layer. This approach has been reported to increase the breakdown strength far beyond the sum of the strengths of individual layers.\cite{schultz_detailed_2013} Schultz et al. studied various combinations of $\mathrm{Al_2O_3}$ (which was grown by anodization) and SU-8 with parylene and were able to achieve better electrical insulation. The same approach is taken here except that there are no parylene or anodized $\mathrm{Al_2O_3}$ used here because this project sought to limit the needed equipment to spin coaters and ovens. It is worth pointing out here that one of the disadvantages of multi-layer dielectric insulation is that increasing the insulation thickness increases the applied voltage to achieve a desired electric field strength. Though a study of thickness optimization will be helpful, we believe this is out of scope for this presented study.

Processing and fabrication details also matter, the electrical properties of PMMA vary with the type of solvent used to dissolve it, the annealing temperature, and the thickness of the layer. Tippo et al. studied the effect of various solvents and concluded that Anisole and dimethylformamide are the best choices as solvents for PMMA.\cite{tippo_effects_2013} Yun et al. studied the effects of the thickness of Anisole based PMMA and concluded that the thickness of a PMMA layer has to be optimized to reduce the number of defects and thus increase the breakdown strength of the PMMA layer.\cite{yun_pentacene_2009} Na \& Rhee reported that a PMMA layer annealed above the glass transition temperature has poor electrical properties as compared to one annealed at lower temperatures.\cite{na_electronic_2006} Zhang et al. studied the electrical properties of PMMA prepared using spin coating and reported the dielectric constant of PMMA to be 3.9 at 1MHZ while the breakdown strength to be 5.8 MV/cm.\cite{zhang_optical_2015}

Over the years, SU-8 has been primarily used in soft lithography to fabricate channels. Because of its excellent chemical, electrical and optical properties, it has a high potential as a dielectric layer. Melai et al. studied the electrical properties of SU-8 of various thicknesses to use it as a dielectric layer and reported the dielectric strength to be $\mathrm{4.43 \pm 0.16 ~ MV/cm}$\cite{melai_electrical_2009} with dielectric constant varying between $3-4$. In real-life applications like electrowetting, Kumar \& Sharma used SU-8 as a hydrophobic, dielectric layer to replace Teflon while achieving the same results.\cite{kumar_su-8_2012} Apart from the excellent electrical properties, SU-8 also exhibits better chemical properties which make them highly etch resistant. Finally, SU-8 also exhibits good optical properties, which makes it ideal for experiments involving optical measurements.\cite{mogensen_integration_2003}

Based on the above-mentioned benefits of the SU-8 and PMMA as dielectric layers, the next section (Material and Channel Fabrication) presents a simplistic yet detailed method to fabricate the insulation layers for electrolytes along with a channel fabricated out of a SU-8 layer. A detailed method is also reported on closing the channel using another SU-8-coated substrate. The results and Discussion section presents the current and voltage characteristics for an ionic solution of various concentrations to understand the insulation properties of the proposed materials. The results show that PMMA and SU-8 are not just very feasible to fabricate, they are also very effective as insulation layers for microfluidic applications that require high electric fields such as dielectrophoresis.

\section{Materials, Channel Fabrication, and Experimental Method}

Two Fluorine doped tin oxide (FTO)-coated glasses with an electrode thickness of $\sim 320nm$ and surface resistivity of $\mathrm{\sim 7 ~ \Omega/cm^2}$ are used as substrates (supplied by MSE Supplies LLC). We shaped the electrodes by masking the substrates using a positive photoresist KL6005 (supplied by Microchem) and etching them. To mask the substrates, KL-6005 was spin-coated at $\mathrm{1000~rpm}$ for 1 minute to get a thickness of 1 micron. Following the coating, the layer was soft-baked at $\mathrm{105^\circ ~C}$ for 3 minutes and then allowed to cool down to room temperature. To mask the desired shape of the electrodes, a photomask (supplied by FineLine Imaging) was used, where the KL6005 layer was exposed to ultraviolet (UV) of wavelength $\mathrm{365~nm}$ using which initiates polymerization. After the exposure, the photoresist was hard-baked at $\mathrm{95^\circ ~ C}$ for three minutes and once again cooled down to room temperature. At last, to remove the exposed part of the photoresist, the substrates were dipped and developed in a Tetramethylammonium hydroxide developer (0.26 N TMAH, supplied by Sigma Aldrich) for 3 minutes. After development, the substrates were cleaned with water and dried with an air gun.

\begin{table}[t]
    \centering
    \begin{tabular}{c|c|c|c}
        Layer &  First step & Second step & Third step\\
            & (rpm) & (rpm) & (rpm)\\
         PMMA & 3000 & 5000 & 100\\
         SU-8 & 3500 & 5000 & 100
    \end{tabular}
    \caption{A multi-speed based spin coating technique for PMMA and SU-8, where the different speeds for three steps are shown.}
    \label{tab:table1}
\end{table}

Upon masking the substrates, Zinc Oxide powder was sprinkled on the slide to cover the whole unmasked area. A wet etch was carried out using $\mathrm{2~M}$ HCL solution spread on the preferred area with a pipette and left for $\mathrm{\sim 3~mins}$ for optimal etching. After that, hot $\mathrm{(\sim 70^\circ ~ C)}$ DI water was poured on it to initiate the cleaning process by removing the zinc powder. Upon that, the slides were wiped with acetone to get rid of any residues of zinc and the photoresist. Inlet and outlet holes were then drilled into one of the substrates. Both substrates were then cleaned in an ultrasonic bath using Acetone, Ethanol, and Isopropyl Alcohol (IPA) for 15 minutes each. Following chemical cleaning, the substrates were dried using an air gun blown over the surface. Finally, to prepare the substrates for spin-coating, they were baked to remove any water at $\mathrm{120^\circ ~ C}$ for 15 minutes. Before starting the coating process, the substrate with holes was covered with black tape on the non-FTO side of the slide to prevent dripping of the photoresist.

\begin{figure}
    \centering
    \includegraphics[width=0.9\textwidth]{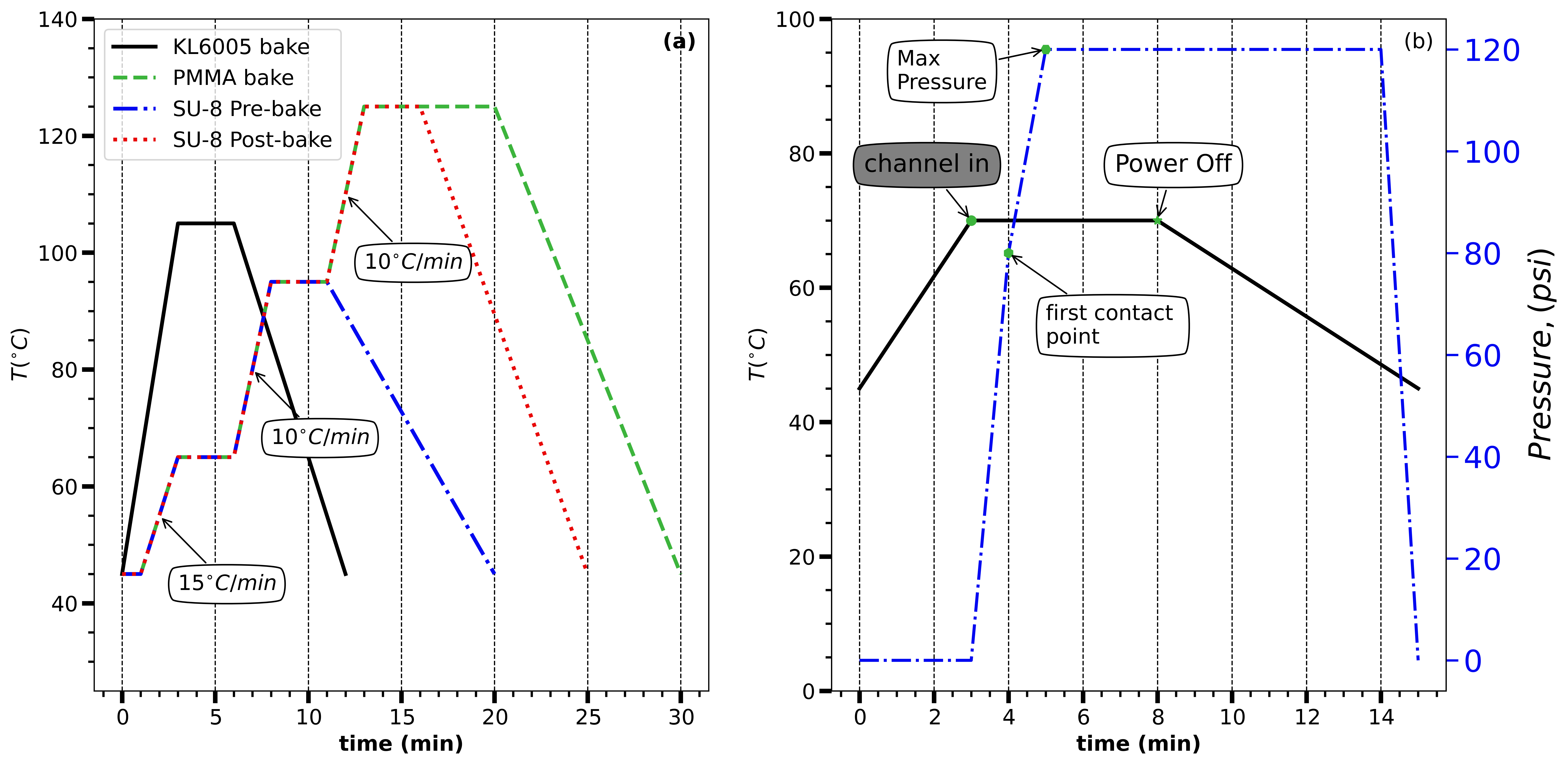}
    \caption{Temperature and pressure profiles. a) Baking temperature profile for different photoresists. (black solid line represents KL-6005, the green dashed line represents PMMA, the blue dash-dot line represents SU-8 prebake, and the red dotted line represents SU-8 post-bake), b) shows the temperature (represented by a black solid line) and pressure (represented by a blue dash-dot line) profile during the SU-8 adhesion of the bottom and top substrates.}
    \label{fig:figure1}
\end{figure}

The first part of the coating starts with spin-coating of two layers of Polymethyl methacrylate (950 PMMA A9, supplied by Microchem Laboratory) to form the base layer of electrode insulation with a thickness of $\mathrm{\sim 1~\mu m}$ each. All photoresists used here are poured on the substrate with a pipette at an angle. This special procedure was adopted to make sure the PMMA covers the whole surface, which reduces the chances of edge bead formation in a rectangular substrate. PMMA was left on the substrate for two minutes to get rid of any air bubbles formed during PMMA pouring. The spin coating parameters are listed in Table \ref{tab:table1}, where a high rpm of 5000 was chosen to ensure a highly dried PMMA layer without edge bead. Following spinning, the baking of PMMA is done on a hot plate where a systematic temperature profile is applied, as shown in Figure \ref{fig:figure1} to avoid the formation of any pinholes. To achieve a smooth surface, the substrate was rotated at regular time intervals in case the hot plate was not level. The same procedure was used to coat the other layer of PMMA on both substrates.

\begin{figure}
    \centering
    \includegraphics[width=0.7\textwidth]{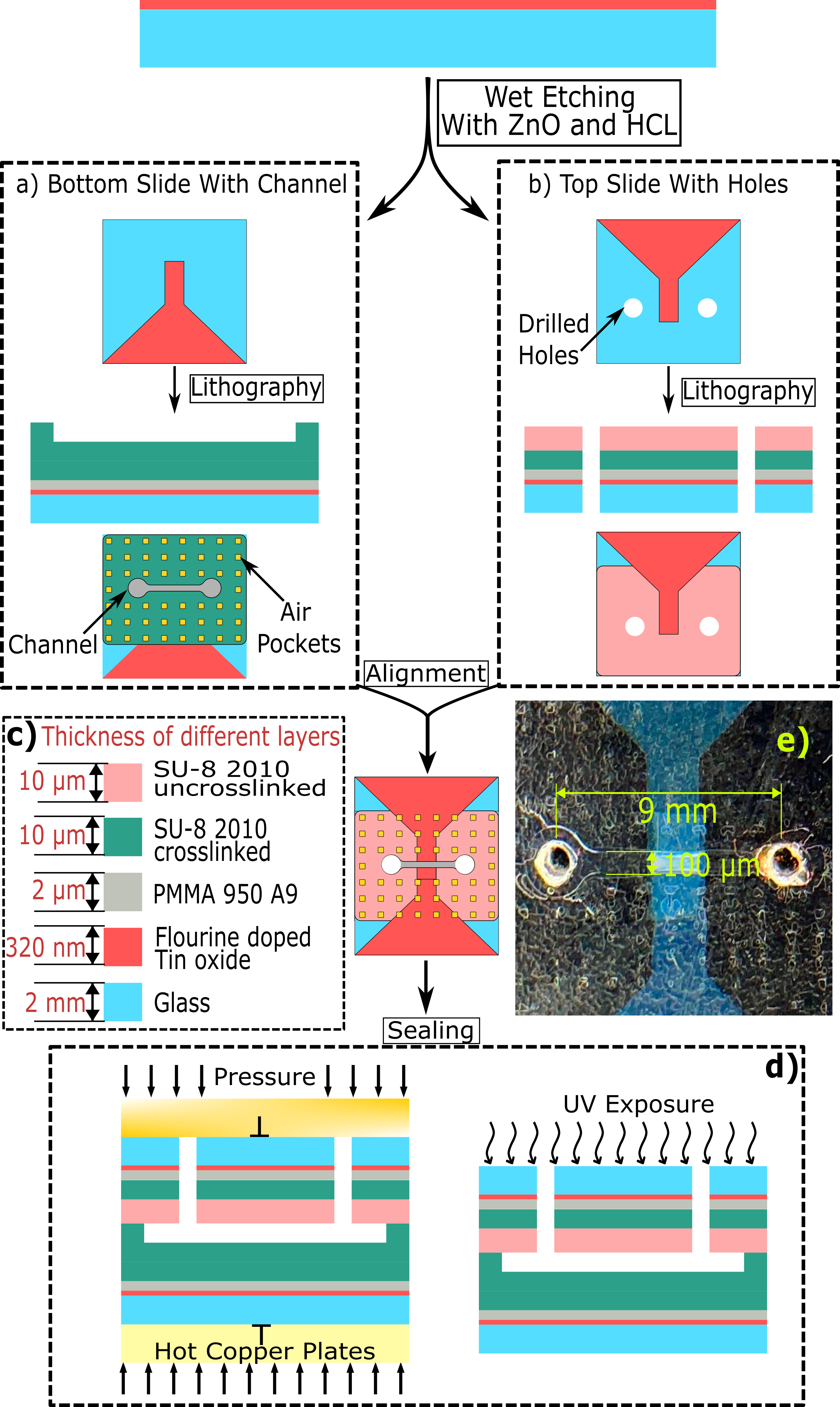}
    \caption{A schematic representing the electrode insulation process and channel fabrication technique. a) top substrate with the channel and the air pockets fabricated on a SU-8 layer during the photolithography process, b) bottom substrate with two drilled holes for the passage of the solution through the channel along with the insulated layers formed through photolithography, c) shows the thicknesses of the different layers,  d) represents the sealing process where pressure and temperature are applied using a hot press to achieve the sealing of two SU-8 layers,  e) shows the channel image.}
    \label{fig:figure2}
\end{figure}

After coating PMMA, a layer of SU-8 2010 (supplied by Microchem Laboratory) is coated on both substrates to form the second layer of electrode insulation. Same procedure as PMMA was used to spread SU-8 and then was left on the substrate for 3 minutes to get rid of any air bubbles formed during SU-8 pouring. The spin coating parameters for SU-8 are listed in Table \ref{tab:table1}, where once again, SU-8 was spun for 3 minutes to minimize the solvent contents and edge beads. After soft-baking according to the temperature profile shown in Figure \ref{fig:figure1}, SU-8 was exposed to 365 nm UV rays to complete the polymerization of monomers and form a solid, hardened layer of SU-8 at the end of hard baking.\cite{SU-8-Technical-data-sheet} A point to be noted here is that the procedure to coat the insulating layers on both the substrates are same until the start of the next steps, as shown in Figure \ref{fig:figure2}. Moving on, the two substrates - one without holes and one with holes - will be referred to as bottom and top substrates respectively, for convenience.

To fabricate the channel on the bottom slide, we coated another layer of SU-8 and soft-baked it as per the temperature profile shown in Figure \ref{fig:figure1}. This SU-8 layer was then exposed to UV using another photomask (supplied by fineline imaging), where the photomask has the channel patterned in black on polyester as shown in Figure \ref{fig:figure2}. The photomask also had small square pockets of area $\mathrm{1*1~ mm^2}$ as shown in Figure \ref{fig:figure2}. The purpose of these small pockets was to allow the SU-8 coated on the top slide to conform and fill them to seal the channel with minimum air gaps.\cite{lima_sacrificial_2015} After exposure and post-bake, unexposed SU-8 was washed by dipping the substrate in a SU-8 developer and then washed with IPA, followed by drying. To achieve a hardened SU-8 layer, the bottom substrate was baked at $\mathrm{125^\circ ~C}$ for 30 minutes and cooled down to room temperature.

Sealing of the channel was achieved by first spin-coating SU-8 on the top substrate, one with holes, and soft-baked using the same temperature profile as shown in Figure \ref{fig:figure1}. The substrate was rotated at regular intervals while softbaking to ensure a smooth surface. After soft bake, both top and bottom slides were aligned and brought in contact, and placed in a temperature-controlled press (manufactured by Dapress) as shown in Figure \ref{fig:figure2}. Before starting the adhesion process, the hydraulic press was set to $\mathrm{70^\circ ~ C}$. This temperature is above the SU-8 uncured glass transition temperature, and pressure ensures the uncured SU-8 will conform to the bottom channel. After the aligned chip is placed in the press, pressure is slowly increased to 120 psi and then maintained for 2 minutes. A detailed temperature and pressure profile for the adhesion process is shown in Figure \ref{fig:figure1} b). The channel was then allowed to cool down to room temperature while maintaining pressure. Once the channel was cooled down, the pressure was released slowly, and the channel was removed from the press. Finally, the entire chip is exposed to 365 nm UV light to permanently bond the two halves, followed by hard bake as shown in Figure \ref{fig:figure2}. 

After sealing the channel, the adhesion test was done using a visual technique to look for any kind of detachment. Figure \ref{fig:figure2} shows clearly that there is a good attachment with an area less than $\mathrm{10~\%}$ is not attached. While all four channels' edges were sealed with UV glue, the insulation layer was removed with a sharp knife and wiped with acetone from the two edges of the channel to make contact with electrodes with electrical wires. The top and bottom FTO-coated glass slides of the sealed channel were attached to an AC power supply via high-voltage electrodes. The holes on the top slides were attached to a micro pump via two tubes (an inlet and another outlet) to supply the solution.

A range of four different concentrations of solutions were prepared to test the insulation strength of the channel. The four different solutions were prepared by dissolving different amounts of sodium chloride (NaCl) salt in deionized water to achieve final NaCl concentrations of $\mathrm{0.1~mM,~ 0.5~mM, ~1.0~mM, ~and ~10~mM}$. The resulting solutions were checked for their conductivity using a conductivity meter (supplied by Apera Instruments). The conductivity for the solutions varied from $\mathrm{0.1~mS/m ~to ~100~mS/m}$ as expected for the chosen concentrations of NaCl solution. The channels were filled with different solutions, and then a sine wave AC voltage was applied to test the insulation strength. For each solution, the peak to peak voltage $\mathrm{(V_pp)}$ was varied from $\mathrm{50~Vpp~ to~ 2000~Vpp}$, while the frequency was varied from $1\mathrm{1 ~kHz ~to ~50~kHz}$. At each set of frequency and voltage, the power was supplied for 5 seconds, while the current and voltage across the channel was recorded using an oscilloscope (PicoScope® 4262, supplied by Pico Technology Ltd.) with a sampling rate of $\mathrm{5~MHz}$. For a set of voltage-current characteristics data, a voltage was fixed, and the frequency was varied between the above-mentioned range and subsequently, the voltage was varied till the channel burnt. The recorded data was then analyzed to estimate the root mean square (RMS) current and voltage values.

\section{Results and Discussions}

\begin{figure}
    \centering
    \includegraphics[width=0.9\textwidth]{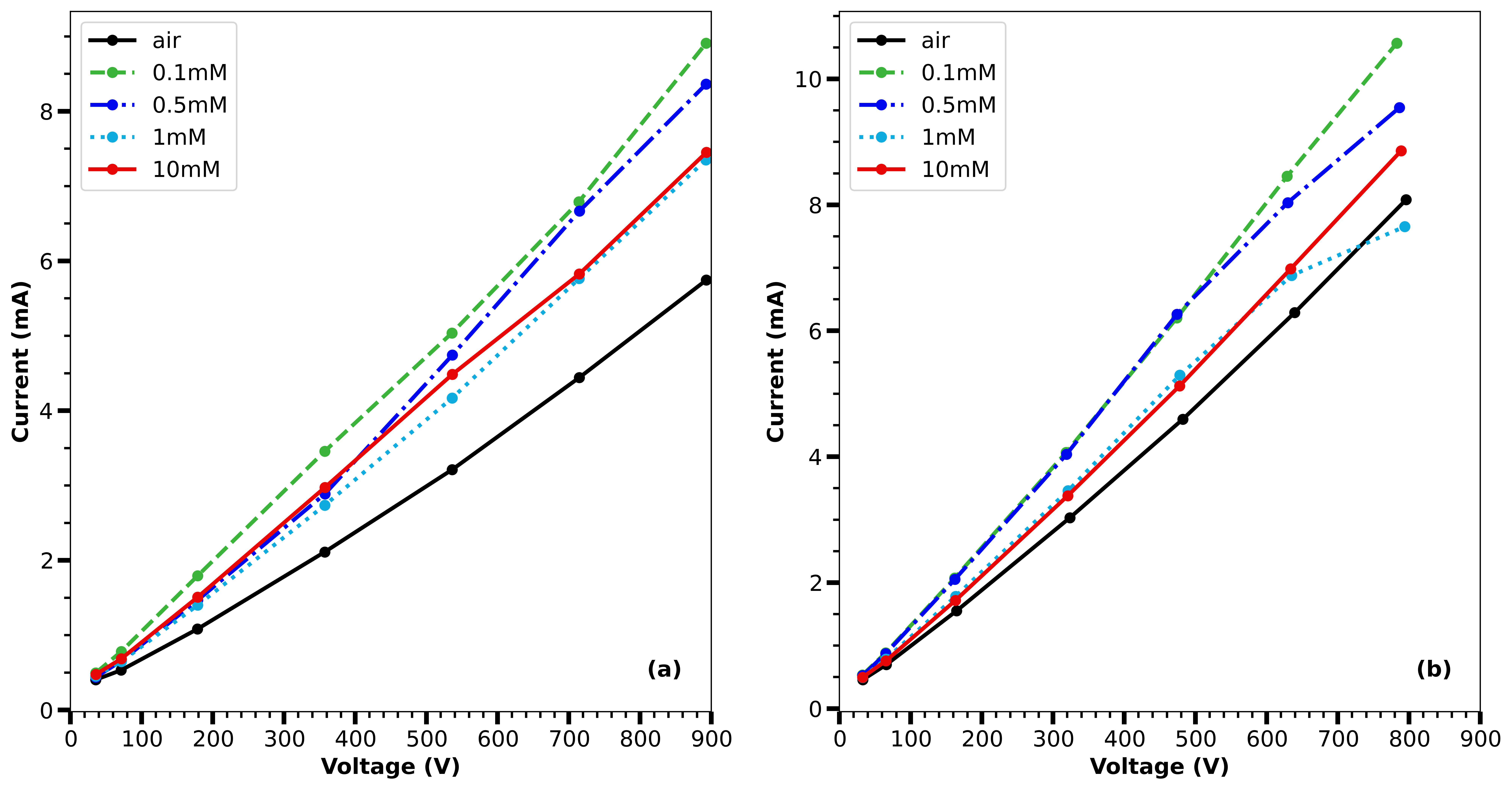}
    \caption{Voltage-Current characteristics of four different solutions (green solid line represents $\mathrm{0.1~mM}$ NaCl solution, blue solid line represents $\mathrm{0.5~mM}$ NaCl solution, cyan solid line represents $\mathrm{1~mM}$ NaCl solution, red solid line represents $\mathrm{10~mM}$ NaCl solution), and air or empty channel (represented by black solid line) for two different frequencies. a) $\mathrm{5~kHz}$, b) $\mathrm{50~kHz}$ ($\bullet$ represents the experimental values).}
    \label{fig:figure3}
\end{figure}

Figure \ref{fig:figure3} a) and b) shows RMS current and voltage values at frequencies $\mathrm{5~kHz~and~50~kHz}$ respectively, for the above-mentioned concentrations of NaCl solutions along with for air that represents empty channel. Figure \ref{fig:figure3} a) and b) also represents the voltage-current (V-I) characteristics at the lowest and highest frequency, i.e., $\mathrm{5~kHz~and~50~kHz}$ applied during the experiment. Though some of the dielectrophoretic applications involve use of very high frequencies (in range of MHz),\cite{sanghavi_electrokinetic_2014} we applied a maximum frequency of $\mathrm{50~kHz}$ to prevent excessive temperature rise of the solution, in case of very conductive solutions like sea water, the temperature can easily rise to the boiling point of the solution due to Joule heating.\cite{anand_effect_2022,anand_effects_2022} Joule heating in electrolytes is a function of applied electric field and frequency, increasing quadratically with the former and linearly with the latter. Since temperature rise due to Joule heating has an established relation, we choose to vary the former and fix the latter to a maximum of $\mathrm{50~kHz}$ to avoid boiling of the electrolytes. Though Joule heating can be a concern in regards to boiling of electrolytes, it poses no threat to the insulation layers because of the very high degradation temperature (above $\mathrm{350^\circ~C}$) of cross-linked SU-8.

In Figure \ref{fig:figure3}, the current is minimum at all applied voltages when there was air inside the channel (solid black lines) due to the absence of ions. The current increases to a maximum at a given voltage as soon as $0.1~mM$ NaCl solution is passed through the channel as shown by green solid lines in Figure \ref{fig:figure3}. While the RMS current value is highest for the low-concentration solution, the current decreases with an increase in concentration and reaches to lowest for the high-concentration solution as shown by red solid lines in Figure \ref{fig:figure3}.

\begin{figure}
    \centering
    \includegraphics[width=0.9\textwidth]{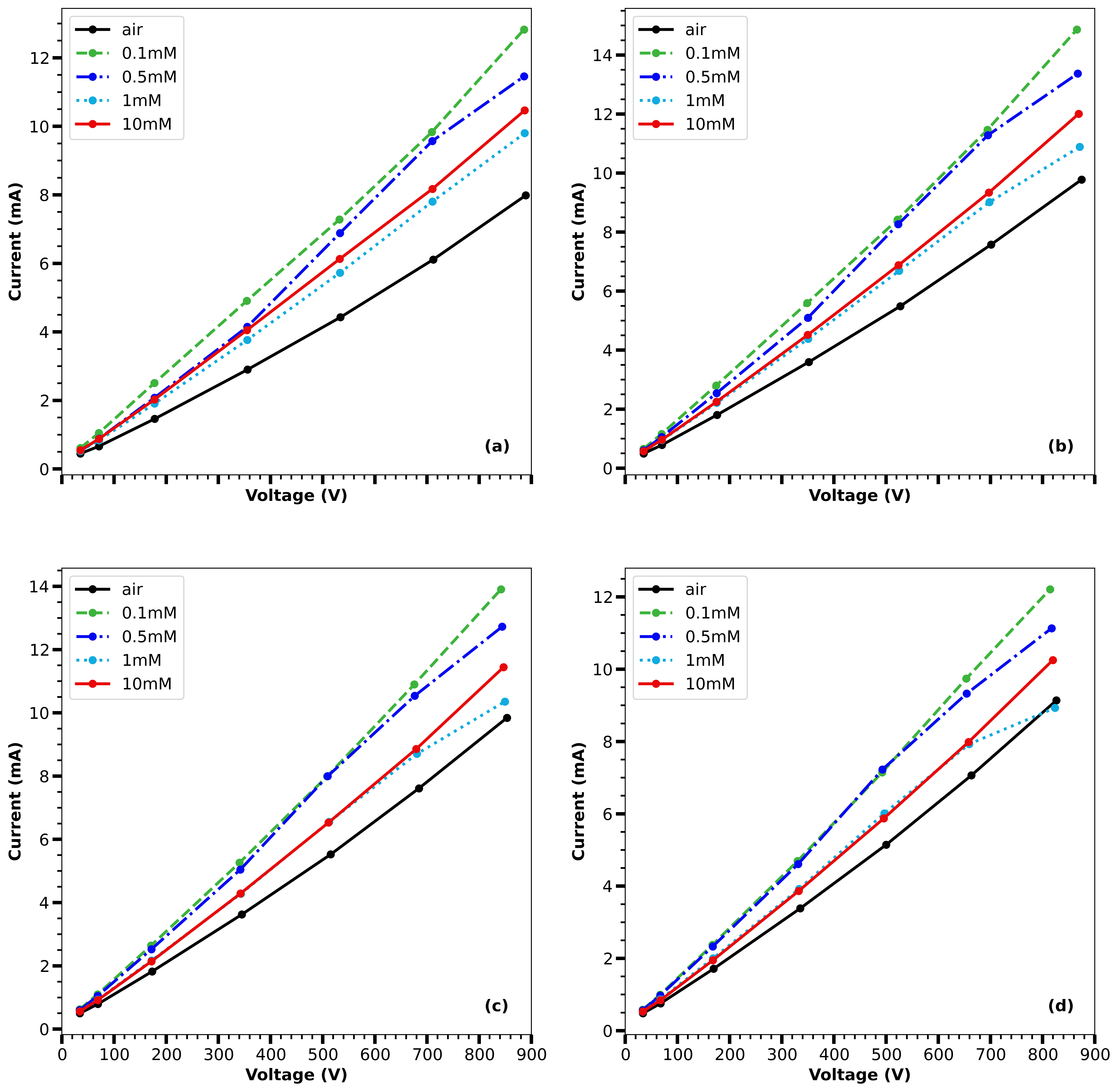}
    \caption{Voltage-Current characteristics of four different solutions (green solid line represents $\mathrm{0.1~mM}$ NaCl solution, blue solid line represents $\mathrm{0.5~mM}$ NaCl solution, cyan solid line represents $\mathrm{1~mM}$ NaCl solution, red solid line represents $\mathrm{10~mM}$ NaCl solution), and air or empty channel (represented by black solid line) for four different frequencies. a) $\mathrm{10~kHz}$, b) $\mathrm{20~kHz}$, c) $\mathrm{30~kHz}$, d) $\mathrm{40~kHz}$ ($\bullet$ represents the experimental values).}
    \label{fig:figure4}
\end{figure}

An established standard way to analyze a microfluidic channel with an electrolyte confined between two blocked electrodes subjected to an AC power supply is to treat it as a series of a capacitor and a resistor.\cite{DF9470100011} Therefore, while it might be expected that an increase in concentration or conductivity should have resulted in an increase in current in Figure \ref{fig:figure3}, the decrease in capacitance due to the decrease in permittivity of the solution increases the electrical impedance. The effect of the increase in the concentration of NaCl on the conductivity and permittivity of the solution has been studied before.\cite{anand_effect_2022} For the above-mentioned frequency range the electrical impedance is more influenced by the permittivity of the medium as the capacitive reactance is two orders of magnitude higher than the conductance. Also, one can observe the effect of the frequency on current while comparing Figure \ref{fig:figure3} a) and b), where for any chosen solution, the current increases with an increase in frequency due to a decrease in capacitive reactance.

The V-I characteristics at frequencies $\mathrm{10 ~kHz, ~20 ~kHz, ~30 ~kHz, ~and ~40 ~kHz}$ are shown in Figure \ref{fig:figure4} a), b), c), and d) respectively. The RMS current and voltage values in Figure \ref{fig:figure4} also have the same characteristics as in Figure \ref{fig:figure3}. In Figures \ref{fig:figure3} and \ref{fig:figure4}, the plots show that at a fixed frequency, the current is proportional to voltage for a range where the voltage strength is low and then it non-linearly varies with the voltage as the voltage strength increases especially for high NaCl concentration solution. In the case of low-concentration solutions, because of the constant electrical impedance, the change in voltage and current is linear in nature. The reason behind the non-linearity in the case of high-concentration solution can be attributed to the formation of an electrical double layer at blocked electrodes, the effects of which are described in detail elsewhere.\cite{anand_effects_2022}

\begin{figure}
    \centering
    \includegraphics[width=0.8\textwidth]{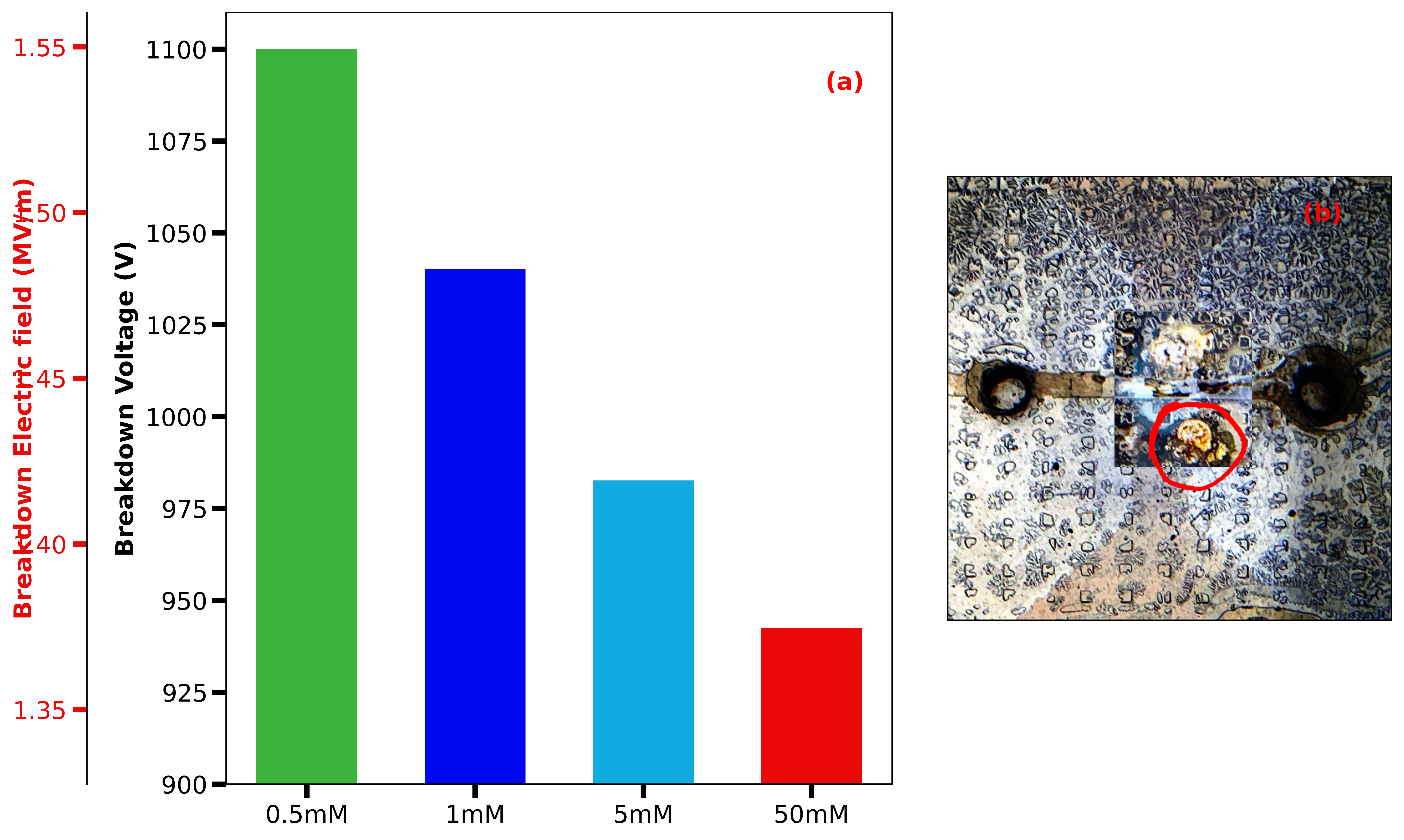}
    \caption{Analysis of electrical insulation breakdown. a) shows the breakdown voltage and corresponding breakdown electric field strength for four different concentrations of NaCl solution (green bar represents $\mathrm{0.1~mM}$ NaCl solution, blue bar represents $\mathrm{0.5~mM}$ NaCl solution, cyan bar represents $\mathrm{1~mM}$ NaCl solution, red bar represents $\mathrm{10~mM}$ NaCl solution) b) shows the image of a channel after the dielectric breakdown (red solid line encircles the failure area).}
    \label{fig:figure5}
\end{figure}

Figure \ref{fig:figure5} a) shows the breakdown voltage and the corresponding breakdown electric field strength for four different concentrations of NaCl solutions. While for all the solutions below $\mathrm{1~mM}$ the breakdown voltage was seen to be $\mathrm{2000~Vpp}$, the breakdown voltage for solutions with concentration $\mathrm{1~mM}$ and higher was $\mathrm{1500~Vpp}$. This shows that as the concentration of ions increases, the breakdown strength of the insulation layers decreases, which is as expected. The high breakdown strength is however, interesting. The breakdown strength for the $\mathrm{0.5 ~mM}$ NaCl solution is $\sim \mathrm{1.55~MV/m}$, and the breakdown strength for $\mathrm{5.0 ~mM}$ NaCl solution decreases to $\sim \mathrm{1.35~MV/m}$.

The failures of the dielectric layer for various solutions at the breakdown voltages are observed to be very similar in nature. The failure image for one of the solutions and the channel is shown in Figure \ref{fig:figure5} b), where the failure area is encircled with a solid red line. The failure of the proposed insulation method at breakdown voltage shows that the SU-8 failed at the edges of electrodes due to bulk breakdown of the dielectric layers when compared to the results produced by Melai et al.,\cite{melai_electrical_2009} where the failure of different thickness of SU-8 dielectric layers was studied.

\section{Conclusions}

A multilayered high dielectric strength electrical insulation fabrication technique is presented here, which has been shown to withstand a voltage as high as $\sim 1000~V$, or equivalently an electric field strength of $\mathrm{\sim1~ MV/m}$ for varying concentrations of NaCl solution. This high dielectric strength was accomplished using a simple insulation fabrication technique and functions well for electrolytes of conductivity as high as $0.1~S/m$, i.e., higher than that of seawater or blood, which is a huge improvement in electrical insulation strength while applying the electric field uniformly over a volume. In addition to providing a better understanding of the dielectric strength of these insulators, these results have the potential to impact several areas of science. For example, dielectrophoretic transport of molecular solutes was recently enabled through the high dielectric strength of these electrodes \cite{anand_dielectric_2023,anand_separations_2023} as was an investigation into the effects of high electric fields on the Raman spectrum of solutions\cite{anand_novel_2023}. Its likely that low-cost electrowetting or other electrohydrodynamics-based applications will also benefit from the demonstrated improvement in dielectric strength.

\section*{Conflicts of interest}
There are no conflicts to declare.

\section*{Acknowledgements}
The authors thank to the donors of the American Chemical Society Petroleum Research Fund for partial support of this research.

\bibliography{main}
\bibliographystyle{iopart-num}

\end{document}